\documentstyle[12pt,amssymb]{article}
\def\greaterthansquiggle{\raise.3ex\hbox{$>$\kern-.75em\lower1ex\hbox{$\sim$}}}
\def\lessthansquiggle{\raise.3ex\hbox{$<$\kern-.75em\lower1ex\hbox{$\sim$}}}
\newcommand{\beq}{\begin{equation}}

\newcommand{\cfeg}{{\em cf., e.g.,}\ }

\newcommand{\eeq}{\end{equation}}
\newcommand{\beqa}{\begin{eqnarray}}
\newcommand{\eeqa}{\end{eqnarray}}
\newcommand{\beqan}{\begin{eqnarray*}}
\newcommand{\eeqan}{\end{eqnarray*}}
\newcommand{\ba}{\begin{array}}
\newcommand{\ea}{\end{array}}

\def\nz{\ifmmode {I\hskip -3pt N} \else {\hbox {$I\hskip -3pt N$}}\fi}
\def\zz{\ifmmode {Z\hskip -4.8pt Z} \else
       {\hbox {$Z\hskip -4.8pt Z$}}\fi}
\def\qz{\ifmmode {Q\hskip -5.0pt\vrule height6.0pt depth 0pt
       \hskip 6pt} \else {\hbox
       {$Q\hskip -5.0pt\vrule height6.0pt depth 0pt\hskip 6pt$}}\fi}
\def\rz{\ifmmode {I\hskip -3pt R} \else {\hbox {$I\hskip -3pt R$}}\fi}
\def\cz{\ifmmode {C\hskip -4.8pt\vrule height5.8pt\hskip 6.3pt} \else
       {\hbox {$C\hskip -4.8pt\vrule height5.8pt\hskip 6.3pt$}}\fi}
\def\au{{\setbox0=\hbox{\lower1.36775ex%
\hbox{''}\kern-.05em}\dp0=.36775ex\hskip0pt\box0}}
\def\ao{{}\kern-.10em\hbox{``}}

\newtheorem{Theorem} {Theorem} [section]

          \newcommand{\N}{{\Bbb N}} 
          \newcommand{\R}{{\Bbb R}}
\newcommand{\Comp}{{\Bbb C}}
{\catcode `\@=11 \global\let\AddToReset=\@addtoreset}
\AddToReset{equation}{section}

\newcommand{\subjclass}[1]{}

\def\scri{\hbox{${\cal J}$\kern -.645em {\raise
      .57ex\hbox{$\scriptscriptstyle (\ $}}}}

\newcommand{\cD}{{\cal D}} 
 
\newcommand{\Si}{\Sigma}

 \newcommand{\eq}[1]{(\ref{#1})}

\newcommand{\commentout}[1]{}

\newcommand{\ee}{\end{equation}} \newcommand{\bea}{\begin{eqnarray}}
\newcommand{\eea}{\end{eqnarray}}
\newcommand{\beaa}{\begin{eqnarray*}}
\newcommand{\eeaa}{\end{eqnarray*}}

\newcommand{\Sext}{\Sigma_{\mathrm ext}}
\newcommand{\Mext}{M_{\mathrm ext}}
\newcommand{\doc}{\langle\langle M_{\mathrm ext}\rangle\rangle}

\begin{document}

\title{On rigidity of analytic black holes}
\author{ Piotr T.\ Chru\'sciel\thanks{ 
On leave of absence from the Institute of
    Mathematics, Polish Academy of Sciences, Warsaw.
  Supported in part by KBN grant \# 2 P301 105 007.
  {\em E--mail}:
    Chrusciel@Univ-Tours.fr} 
\\ D\'epartement de Math\'ematiques\\Facult\'e
des Sciences\\ Parc de Grandmont\\ F37200 Tours, France}

\maketitle

\begin{abstract} 
 We establish global extendibility (to the domain of outer
 communications) of locally defined isometries of appropriately regular
 analytic black holes. This allows us to fill  a gap in the
 Hawking--Ellis proof of  black--hole rigidity,  for
 ``non--degenerate'' black--holes. 
\end{abstract}

\section{Introduction}

According to Hawking and Ellis \cite[Prop. 9.3.6]{HE}, under
appropriate conditions, which include analyticity of all the objects
under consideration, the event horizon of a stationary, say
electro--vacuum, black hole space--time $(M,g)$ is necessarily a
Killing horizon. More precisely, the isometry group of $(M,g)$ should
contain an $\R$ subgroup, the orbits of which are tangent to the black
hole horizon. In order to substantiate their claim the authors of
\cite{HE} first argue that for each $t$ the map defined as the
translation by $t$ along the appropriately parameterized generators of
the event horizon extends to an isometry $ \phi_t$ in a neighborhood
of the event horizon.  Next they assert that for all $t$ one can
analytically continue $ \phi_t$ to the whole space--time, to obtain a
globally defined one parameter group of of isometries.  This last
claim is wrong, which\footnote{The construction that follows is a
  straightforward adaptation to the problem at hand of a
  construction in \cite[Section 5]{ChRendall}.} can be seen as
follows: Let $( M, g_{ab})$ be the extension of the exterior region of
the Kerr space--time consisting of ``two type I regions and two type
II regions'', as described in Section 5.6 of \cite{HE} (thus $( M,
g_{ab})$ consists of the four uppermost blocks of Figure 28, p. 165 in
\cite{HE}). Let $\phi_t$ denote those isometries of $( M, g_{ab})$
which are time--translations in an asymptotic region $\Mext$, and let
$\doc$ denote the domain of outer communication of $( M, g_{ab})$ as
determined by $\Mext$ ({\em cf.\/} eq. \eq{mext} below; $M_{ext}$
corresponds to one of the blocks ``I'' of Figure 28 of \cite{HE}).
Let $\Si$ be any asymptotically flat Cauchy surface of $( M, g_{ab})$
(thus $\Si$ has two asymptotic regions), and let ${\cal E}$ be any
embedded two-sided three-dimensional sub-manifold of
$\mbox{int}\cD^+(\Si;M)\setminus\doc$, invariant under $\phi_t$.  We
shall moreover suppose that $M\setminus\overline{{\cal E}}$ is
connected, and that $\cal E$ is {\em not} invariant under the $U(1)$
factor of the isometry group of $( M, g_{ab})$.  Let $(M_a,g_a)$,
$a=1,2$, be two copies of $M\setminus\overline{ {\cal E}}$ with the
metric induced from $g$.  As ${\cal E}$ is two-sided, there exists an
open neighborhood ${\cal O}$ of ${\cal E}$ such that ${\cal E}$
separates ${\cal O}$ into two disjoint open sets ${\cal O}_a$,
$a=1,2$, with $\overline{{\cal O}_1}\cap \overline{{\cal
    O}_2}=\overline{ {\cal E}}$, ${\cal O}_1\cap {\cal
  O}_2=\emptyset$. Let $\psi _a$ denote the natural embedding of
${\cal O}_a$ into $M_a$. Let $M_3$ be the disjoint union of $M_1,M_2$
and ${\cal O}$, with the following identifications: a point $p\in{\cal
  O}_a\subset{\cal O}$ is identified with $\psi _a(p)\in M_a$. It is
easily seen that $M_3$ so defined is a Hausdorff topological space.
 
We can equip $M_3$ with the obvious real analytic manifold structure
and an obvious  metric $g_3$ coming from $(M_1,g_1)$, $(M_2,g_2)$ and
$({\cal O},g|_{\cal O})$. Note that 
$g_3$ is real analytic with respect to this structure. Let finally
$(M_4,g_4)$ be any   
maximal\,\footnote{\label{fnmaximal} \cfeg
    \cite[Appendix C]{SCC} for a proof of existence of space--times
    maximal with respect to some property. It should be pointed out
that there is an error in that proof, as the relation $\prec$ defined
there is not a partial order. This is however easily corrected by
adding the requirement that the isometry $\Phi$ considered there
restricted to some fixed three--dimensional hypersurface be the
identity.} vacuum real analytic extension of $(M_3,g_3)$.  
Then $(M_4,g_4)$ is a maximal vacuum  real analytic extension of  
$\doc$ which clearly is not isometric to $(M,g)$. 

The space--time $(M_4,g_4)$ satisfies all the hypotheses
of \cite{HE}.  The connected component of the identity of the group of
isometries drops down from  ${\R}\times U(1)$ (in the case of $( M,
g_{ab})$) to $\R$ (in the case of $(M_4,g_4)$), as
all the orbits of the rotation group acting on $( M, g_{ab})$ meeting
${\cal E}$ are incomplete in  $(M_4,g_4)$. 

Topological games put aside, the method of proof suggested in
\cite{HE} of analytically extending $ \phi_t$ faces the problem that $
\phi_t$ might potentially be analytically extendible to a proper
subset\footnote{ In the physics literature there seem to be
  misconceptions about existence and uniqueness of analytic extensions
  of various objects.  As a useful example the reader might wish to
  consider the (both real and complex) analytic function $f$ from,
  say, the open disc $D(1,1/2)$ of radius $1/2$ centered at $1$ into
  $\Comp$, defined as the restriction of the principal branch of $\log
  z$. Then: 1) There exists no analytic extension of $f$ from
  $D(1,1/2)$ to $\Comp$. 2) There exists no unique maximal subset of
  $\Comp$ on which an analytic extension of $f$ is defined.} of the
space--time only. One can nevertheless hope that the analyticity of
the domain of outer communication and some further conditions, as {\em
  e.g.\/} global hyperbolicity thereof, allow one to extend the
locally defined isometries at least to the whole domain of outer
communications.  The aim of this paper is to show that this is indeed
the case.  More precisely, we wish to show the following:

\begin{Theorem}\label{T1}
Consider an   analytic 
space--time $(M,g_{ab})$ with a Killing vector field
$X$ with complete orbits. Suppose that $M$ contains an asymptotically 
flat three--end $\Sigma_{\mathrm ext}$ with time-like ADM 
four--momentum, and with 
 $X(p)$ --- time-like
for  $p\in\Sigma_{\mathrm ext}$. (Here asymptotic flatness is defined
in the sense of eq.\ \eq{K.99} with $\alpha > 1/2$ and
$k\ge 3$.)
 Let $\langle\langle
M_{ext}\rangle\rangle$ denote 
the domain of outer communications associated with 
$\Sext$ as defined below,  assume that $\doc$ is globally hyperbolic and
simply connected. If there exists a Killing vector field $Y$, which is not
a constant multiple of $X$, defined
on an open subset $\cal O$ of $\doc$, then the isometry group
of $\doc$ (with the metric obtained from $(M,g_{ab})$ by restriction)
contains $\R \times U(1)$.
\end{Theorem}

{\bf Remarks} 
\begin{enumerate}

\item It should be noted that no field equations or energy
  inequalities are assumed.
        
\item Simple connectedness of the domain of outer communications
  necessarily holds when a positivity condition is imposed on the
  Einstein tensor of $g_{ab}$ \cite{galloway-topology}.\footnote{{\em
      cf.\/} also \cite{ChWald,Jacobson:venkatarami,HE} for similar
    but weaker results. Note that in the stationary black hole
    context, under suitable hypotheses one can use Theorem~\ref{T2}
    below to obtain completeness of orbits of $X$ in $\doc$, and then
    use \cite{ChWald} to obtain simple--connectedness of $\doc$.}

\item When a positivity condition is imposed on the Einstein tensor of
  $g_{ab}$, the hypothesis of time-likeness of the ADM momentum can be
  replaced by that of existence of an appropriately regular Cauchy
  surface in $(M,g_{ab})$.  See, {\em e.g.}, \cite{Horowitz} and
  references therein; {\em cf.\/} also \cite{ChBeig1} for a recent
  discussion.
        
\item It should be emphasized that no claims about isometries of
  $M\setminus \doc$ (with the obvious metric) are made.
\end{enumerate}

Theorem \ref{T1} allows one to give a corrected version of the rigidity
theorem, the reader is referred to \cite{ChAscona} for a precise
statement together with a proof.

It seems of interest to remove the condition of completeness of the
Killing orbits of $X$ above. Recall that completeness of those
necessarily holds \cite{Chorbits} in maximal globally hyperbolic, say
vacuum, space--times under various conditions on the Cauchy data.  (It
was mentioned in \cite{Chnohair} that the results of \cite{Chorbits}
generalize to the electro--vacuum case.) Those conditions are,
however, somewhat unsatisfactory in the black hole context for the
following reasons: recall that the existing theory of uniqueness of
black holes gives only a classification of domains of outer
communication $\doc$. Thus in this context one would like to have
results which do not make any hypotheses about the global properties
of the complement of $\doc$ in $M$.
  Moreover the hypotheses of those results of \cite{Chorbits} which
  apply when degenerate Killing horizons are present require further
  justification. Here we wish to raise the question, whether or not it
  makes sense to talk about a stationary black hole space--time for
  space--times for which the Killing orbits are not complete in the
  asymptotic region.  We do not know an answer to that question.  It
  is nevertheless tempting to decree that in ``physically reasonable"
  stationary black hole space--times the orbits of the Killing vector
  field $X$ which is time-like in the asymptotically flat three--end
  $\Sext$ are complete through points in the asymptotic region
  $\Sext$. One would then like to be able to derive various desirable
  global properties of $\doc$ using this assumption.  Our second
  result in this paper is the proof that in globally hyperbolic
  domains of outer communication the orbits of those Killing vector
  fields which are time-like in $\Sext$ are complete ``if and only if"
  they are so\footnote{The quotation marks here are due to the fact
    that in our approach the asymptotic four--end $\doc$ is not even
    defined when the orbits of $X$ through $\Sext$ are not complete.
    In that last case one could make sense of this sentence using
    Carter's definition of the domain of outer communication
  \cite{CarterlesHouches}, 
    involving Scri.}  for points $p\in\Sext$ (it should be emphasized that,
  in contradistinction to \cite{Chorbits}, no maximality hypotheses
  are made and no field equations are assumed below; similarly no
  analyticity or simple connectedness conditions are made here):
        
\begin{Theorem}\label{T2}
  Consider a space--time $(M,g_{ab})$ with a Killing vector field $X$
  and suppose that $M$ contains an asymptotically flat three--end
  $\Sigma_{\mathrm ext}$, with $X$  time-like in $\Sigma_{\mathrm
    ext}$.  (Here the metric is assumed to be twice differentiable,
  while asymptotic flatness is defined in the sense of eq.\ \eq{K.99}
  with $\alpha > 0$ and $k\ge 0$.)  Suppose that the orbits of $X$
  are complete through all points $p\in \Sext$. Let $\langle\langle
  M_{\mathrm ext}\rangle\rangle$ denote the domain of outer communications
  associated with $\Sext$ as defined below. If $\doc$ is globally
  hyperbolic, then the orbits of $X$ through points $p\in \doc$ are
  complete.
\end{Theorem}

In view of the recent classification of orbits of Killing vector field
in asymptotically flat space--times of \cite{ChBeig2} it is of
interest to prove the equivalent of Theorem~\ref{T2} for
``stationary--rotating'' Killing vectors $X$, as defined in
\cite{ChBeig2}.
In Theorem~\ref{Tnowy} below we prove that generalization.

 \section{Definitions, proof of Theorem {\protect\ref{T1}}.}
 \label{proof1}
 
 Throughout this work all objects under consideration are assumed to be smooth.
 For a  vector field $W$ we denote by $\phi_t[W]$ the (perhaps 
 defined only locally) flow generated by W. Consider a Killing
 vector field $X$ which is time-like for $p\in\Sext$. If the orbits 
 $\gamma_p$ of $X$ are complete through points  $p\in\Sext$, then
we define the asymptotically flat four--end $\Mext $ by
\begin{equation}
\label{mext}
 \Mext = \cup_{t\in \R} \phi_t[X](\Sext),
\end{equation}
and the {\em domain of outer communications\/} $\doc$ by
$$ \doc = J^-(\Mext)\cap J^+(\Mext).$$

 Let $R > 0$ and let $(g_{ij},K_{ij})$ be 
initial data on $\Sext\equiv\Sigma_R \equiv {\Bbb R}^3 \setminus B(R)$
satisfying 
\beq
g_{ij} - \delta_{ij} = O_{k} (r^{-\alpha}), \qquad
 K_{ij} = O_{k-1} (r^{-1-\alpha}), 
\label{K.99}
\eeq
with some $
k\ge 1$ and some $0<\alpha< 1$. A set $(\Sext,g_{ij},K_{ij})$ satisfying
the above will be called {\em an asymptotically flat three--end}. 
Here a function $f$, defined on
$
\Sigma_R
$,
is said to be $ O_k(r^\beta)$ if there exists a constant $C$ such that
we have
$$
0 \leq i \leq k \qquad |\partial^i f| \leq C r^{\beta - i}.
$$

We shall need the following result, which is a straightforward
consequence\footnote{Actually in  \cite{Nomizu} it is assumed that 
 $(M,g_{ab})$ is Riemannian. The reader will note that all
 the assertions and proofs of \cite{Nomizu} remain valid word for word
 when ``Riemannian" is replaced by ``pseudo--Riemannian".}
  of what has been proved in \cite{Nomizu}:
  
\begin{Theorem}[Nomizu]\label{TNomizu}
Let $(M,g_{ab})$ be a (connected) simply connected analytic pseudo--Riemannian 
manifold, and suppose that there exists a Killing vector field $Y$ defined
on an open connected subset $\cal O$ of M. Then there exists a Killing 
vector field $\hat Y$ defined on $M$ which 
coincides with $Y$ on $\cal O$.
\end{Theorem}

Let us pass to the proof of Theorem~\ref{T1}. Without loss of generality
we may assume that $X$ is future oriented for $p\in\Sext$.
Simple connectedness and analyticity of $\doc$ together with Theorem
\ref{TNomizu} allow us to conclude that 
the Killing vector $Y$ can be globally extended to a Killing vector field
$\hat Y$ defined on $\doc$. The time-likeness of the ADM four--momentum 
$p^\mu$ allows us to use  the results in \cite{ChBeig2} to assert that
there  exists a linear combination $Z$ (with constant coefficients) 
of $X$ and $\hat Y$ which has complete periodic orbits through all
points $p$ in $\Mext$ which satisfy  $r(p)\ge R$, for 
some $R$. (Moreover $Z$ and $X$ commute.) To prove Theorem~\ref{T1} we need 
to show that the orbits of $Z$ are complete (and periodic) for
all $p\in \doc$.

Consider, thus, a point $p\in \doc$. There exist $q_\pm \in \Mext$, with
$r(q_\pm)\ge R$, such that $p\in J^- ( q_+)\cap J^+ ( q_-)$. Completeness
 and periodicity of the orbits
$\gamma_{q_\pm}[Z]\equiv \cup_{t\in\R}\phi_t[Z](q_\pm)$
of $Z$ through  $q_\pm$ implies that the sets $\gamma_{q_\pm}[Z]$
 are compact.
Global hyperbolicity of $\doc$ implies then that 
$$ K\equiv J^- (\gamma_{q_+}[Z])\cap J^+ (\gamma_{q_-}[Z])$$
 is compact.
 
For $q\in \doc$ let $t_\pm(q)\in \R \cup \{\pm \infty\}$ be 
the forward and backward life time of orbits of $Z$ 
through $q$, defined by the requirement that $(t_-(q),t_+(q))$ is the 
largest
connected interval containing $0$ such that the solution $\phi_t[Z](q)$ 
of the equation $d\phi_t[Z](q)/dt = \circ \phi_t[Z](q)$ is defined for
all $t\in (t_-(q),t_+(q))$. From continuous dependence of solutions of 
ODE`s upon initial values it follows that $t_+$ is a lower 
semi--continuous function and $t_-$ is an upper
semi--continuous function.

Let $\gamma:[0,1]\rightarrow M$ be any future oriented causal curve 
such that $\gamma(0)=q_-$, 
$\gamma(1)=q_+$,
and $p\in \gamma$. Set
\begin{equation}
\label{Tdef}
T_+ = \inf _{q\in \gamma} t_+(q), \qquad 
T_- = \sup _{q\in \gamma} t_-(q).
\end{equation}
Here and elsewhere 
$\inf$ and $\sup$ are taken in $\R \cup \{\pm \infty\}$. If $T_\pm = \pm 
\infty$ we are done, suppose thus that $T_+\ne \infty$; the case 
$T_-\ne -\infty$ is analyzed in a similar way.
By lower semi--continuity of $ t_+$ and compactness of $\gamma$
there exists $\tilde p\in \gamma$ such that $t_+(\tilde p)= T_+$. By
global hyperbolicity the
family of causal curves $\phi_t[Z](\gamma)$, $t\in [0,T_+)$,
accumulates at a causal curve $\tilde\gamma\subset K$. Consequently the 
orbit $ \phi_t[Z](\tilde p)$, $t\in [0,T_+)$, has an accumulation point
in $K$. It follows that $\phi_t[Z](\tilde p)$ can be extended beyond
$T_+$, which gives a contradiction  unless
$T_+=\infty$, and the result follows.\hfill $\Box$

\section{Proof of Theorem {\protect\ref{T2}}.}
\label{proof2}

{\bf Proof of Theorem {\protect\ref{T2}:}} Without loss of generality
we may suppose 
that $X$ is future oriented for $p\in \Sext$. 
 Consider a point $p\in\doc$, there exist
$p_\pm\in \Mext$ such that $p\in J^+ (p_-)\cap J^-(p_+)$. Let $ \Sigma$ be 
a Cauchy surface for $\doc$, without loss of generality we may assume
that $p_-\in I^- (\Sigma)$ and $p_+\in I^+ (\Sigma)$.
Let $t_\pm$ be defined as in the proof of Theorem~\ref{T1}, we have
$t_-(p_\pm)=-\infty$, $t_+(p_\pm)=\infty$. Let $\gamma:[0,1]\rightarrow 
\doc$ be any causal curve such that $\gamma (0)=p_-$, $\gamma (1)=p_+$,
and $p\in\gamma$. Define $T_\pm$ by eq.\ \eq{Tdef}. By lower 
semi--continuity of $t_+$ there exists $\tilde p\in \gamma$ such
that $t_+(\tilde p)=T_+$. Define
$$
\tilde \Omega = \{ s \in [0, T_+):   \phi_s[X](\tilde p)\in 
I^-(\Sigma)\}\ .
$$
Consider any $s\in\tilde\Omega$. Then the  curve obtained 
by concatenating $\phi_t[X]( p_-)$, $t\in [0,s]$, with
 $\phi_s[X](\gamma)$ is a future directed causal curve which starts at
 $p_-$ and passes through $\phi_s[X](\tilde p)$, hence
 \begin{equation}
        s\in\tilde\Omega \quad \Rightarrow \quad \phi_s[X](\tilde p) \in K \equiv
        J^+(p_-)\cap J^-(\Sigma)\ .
        \label{compact}
 \end{equation}
 By global hyperbolicity of $\doc$ the set $K$ is compact.  If
  $\tilde\Omega= \emptyset$ set $\omega =0$, otherwise set
 $$
 \omega = \sup \tilde \Omega\ .
 $$
 Consider any sequence $\omega_i\in \tilde \Omega$ such that
 $\omega_i\rightarrow \omega$. By \eq{compact} and by compactness of $K$ 
 the sequence $\phi_{\omega_i}[X](\tilde p)$ has an accumulation point in 
 $K$. It follows that $\omega < T_+$. 
 
 By definition of $\omega$ we have $ \phi_s[X](\tilde p)\in 
J^+(\Sigma)$ for all $\omega \le s < T_+$. By Lemma 2.5 of \cite{Chorbits}
it follows that $T_+=\infty$. As $t_+(p)\ge t_+(\tilde p) =T_+$ we obtain
 $t_+(p)=\infty$. The equality $t_-(p)=-\infty$ for all $p\in\doc$ is 
 obtained
 similarly by using the time--dual version of Lemma 2.5 of \cite{Chorbits}.
 \hfill $\Box$

Before presenting a generalization of Theorem~\ref{T2} which covers
the case of ``stationary--rotating'' Killing vectors, as defined
in\cite{ChWald,ChBeig2}, we need to introduce some
terminology. Following \cite{ChWald} we shall say that the orbit
through $p$ of a Killing vector field $Z$ is time--oriented if there
exists $t_p>0$ such that $\phi_{t_p}[Z](p)\in I^+(p)$. It then follows
that for all $ \alpha \in\ R$ and all $ z\in\N$ we have
$\phi_{\alpha+zt_p}[Z](p)\in I^+(\phi_{\alpha}[Z](p))$: if 
$\gamma$ is a timelike curve from $p$ to $\phi_{t_p}[Z](p)$, one
obtains a timelike curve from $\phi_{\alpha}[Z](p)$ to   
$\phi_{\alpha+zt_p}[Z](p)$ by concatenating $\phi_{\alpha}[Z](\gamma)$
with $\phi_{\alpha+t_p}[Z](\gamma)$ with
$\phi_{\alpha+2t_p}[Z](\gamma)$, etc.

A trivial example of a Killing vector field with time--oriented orbits
is given by a timelike Killing vector field. A more interesting example
is that of ``stationary--rotating'' Killing vector fields, as
considered in \cite{ChWald,ChBeig2} ---  loosely speaking, those
are Killing vectors which behave like $\alpha \partial/\partial t +
\beta \partial/\partial \phi$ in the asymptotic region, with $\alpha$
and $\beta$ non--vanishing, where $\phi$ is an angular
coordinate. Thus the theorem that follows 
applies in the ``stationary--rotating'' case.

\begin{Theorem}\label{Tnowy}
  The conclusion of Theorem~\ref{T2} will hold if to its hypotheses
  one adds the requirement that $k$ in \eq{K.99} is
  larger than or equal to 2, and if the hypothesis that $X$ is
  timelike is replaced by the assumption that the orbits of $X$
are 
{\em \/time--oriented} through all $p\in\Sext$.
\end{Theorem}

{\bf Proof:} 
The proof
is achieved by a minor modification of the proof of Theorem~\ref{T2},
as follows: Let $p_\pm$ be as in that proof, from the asymptotic
behavior of Killing vector fields in asymptotically flat space--times
({\em cf.\/ e.g.\/} Section 2 of \cite{ChBeig1}) it follows that we
can without loss of generality assume that
\begin{eqnarray*}
&\phi_{2\pi}[X](p_+)\in I^+(p_+),\qquad \phi_{2\pi}[X](p_-)\in
I^+(p_-)\ ,&
\\
& \forall s\in [0,2\pi]\quad \phi_{s}[X](p_-)\in
I^-(\Si)\ ,\quad
  \phi_{s}[X](p_+)\in
I^+(\Si)\ .
&
\end{eqnarray*}
The proof proceeds then as before, up to the definition of the set
$K$, eq. \eq{compact}. In the present case that definition is replaced
by
$$
  K \equiv
        J^+(\cup_{s\in[0,2\pi]}\phi_s[X](p_-))\cap J^-(\Sigma)\ .
$$
This set is again compact, in view of global hyperbolicity of
$\Mext$. The fact that for $s\in\tilde\Omega$ we have
$\phi_s[X](\tilde p)\in K$ follows by considering the causal curve
obtained by concatenating a causal curve $\gamma_1$ from
$\phi_{s-\lfloor s/2\pi\rfloor2\pi}[X](\tilde p)$  to
$\phi_s[X](\tilde p)$ with $\phi_s[X](\gamma)$. Here $\lfloor \alpha
\rfloor$ denotes the largest 
integer smaller than or equal to $\alpha$;  the existence of $\gamma_1$
is guaranteed  by our discussion above.  \hfill $\Box$

{\bf Acknowledgments:} The author is grateful to I. R\'acz for
comments about a previous version of this paper.

\bibliography{/users/piotr/prace/references/hip_bib,%
/users/piotr/prace/references/reffile,%
/users/piotr/prace/references/newbiblio} \bibliographystyle{amsplain}
\end{document}